\documentclass[preprint,amsmath,amssymb]{revtex4}
\usepackage{graphicx}
\usepackage{dcolumn}
\usepackage{bm}
\begin{document}
\title{On Franklin's relativistic rotational transformation and its modification}
\author{ M.~Nouri-Zonoz  \footnote{Electronic address:~nouri@ut.ac.ir, corresponding author}
, H.~Ramezani-Aval  \footnote{Electronic
address:~hramezania@ut.ac.ir} and R.~Gharechahi\footnote{Electronic
address:~r.gharechahi@ut.ac.ir}} 
\address {Department of Physics, University of Tehran, North Karegar Ave., Tehran 14395-547, Iran.}
\begin{abstract}
Unlike the Lorentz transformation which replaces the Galilean transformation among inertial frames at high relative velocities, 
there seems to be no such a consensus in the case of coordinate transformation between inertial frames and uniformly rotating ones.
There have been some attempts to generalize the Galilean rotational transformation 
to  high rotational velocities. Here we introduce a modified version 
of one of these transformations proposed by Philip Franklin in 1922. The modified version
is shown to resolve some of the drawbacks of the Franklin transformation, specially with respect to the 
corresponding spacetime metric in the rotating frame.
This new transformation introduces non-inertial eccentric observers on a uniformly rotating disk and the corresponding metric 
in the rotating frame is shown to be consistent with the one obtained through Galilean rotational transformation for points close to 
the rotation axis. Employing the threading formulation of spacetime decomposition, spatial distances and time intervals in the spacetime metric of a rotating observer's frame are also discussed.
\end{abstract}
\maketitle
\section{Introduction}
{\it ``There is no relativity of rotation''}. This relatively famous quote by Feynman \cite{Feynman} may look as the final word on the discussion of rotation in the context of special relativity. Based on the fact that the presence of acceleration in a uniformly rotating frame, by the equivalence principle, takes us into the realm of general relativity may convince one not to bother with the formulation of rotation in the context of special relativity and look for the resolution of each rotation-based problem in general relativity and in the suitably chosen/constructed solutions of Einstein field equations (which are of course not usually available). Indeed the problem of the relativistic rigidly rotating disk and the spacetime metric in such a frame has been claimed to be the missing link that led Einstein to the introduction of inevitable relation between curved spacetimes and gravitational fields in the years between 1912 to 1913 \cite{Stach}. On the other hand rotation and rotating frames have always been a source of confusion while treated in the context of special relativity; the famous example is the {\it Ehrenfest's Paradox} \cite{Ehren}. Indeed, looking at the literature \cite{Rel}, one finds how diverse are ideas on the relativistic physics in rotating frames and consequently how distant we are from establishing a general consensus even over the main concepts and notions in this subject \footnote{We refer the reader to the preface of \cite{Rel} by J. Stachel and also the detailed historical survey by \O{}. Gr\o{}n in the same reference.}. So in practice one uses either the Galilean rotational transformation (GRT), which is only valid for centrally rotating observers, or consecutive Lorentz transformations between an inertial (laboratory) frame and comoving inertial frames which are  momentarily at rest with respect to the non-inertial rotating observers (eccentric observers). The latter could be obtained either by employing the so-called  {\it hypothesis of locality} along with the same procedure which led to the Fermi coordinates of an accelerated spinning observer \cite{Mashhoon}, or by reducing a general Lorentz transformation obtained for accelerated spinning frames \cite{Nelson} to the case of rotating frames \cite{Nikolic}. Another alternative is the introduction of a relativistic rotational transformation (RRT) which is the main subject of the present paper. \\
It seems that Ehrenfest's Paradox is a good starting point to begin our discussion on rotation and RRTs. To explain the Paradox we consider two frames/observers one at rest (the laboratory observer/frame) and the other one rotating counter-clockswise around it  with constant angular velocity $\Omega$ (the rotating observer/frame) measured by/in the inertial (non-rotating) observer/frame. At this point we use frames (set of clocks and extended fiduciary triad axes) and observers interchangeably but to be more precise one should differentiate between them, for a rotating frame is a non-inertial frame but not all observers in a rotating frame are non-inertial. In other words we should distinguish between a centrally rotating observer (i.e. at the center of the disk) which is an inertial observer and those at nonzero radii which are non-inertial. We will elaborate on this point later in this section. Using cylindrical coordinates we denote the spacetime points in the non-rotating frame with coordinates $(t, r, \phi , z)$ and in the one rotating around the $z$($z^\prime$)-axis with $(t^\prime, r^\prime, \phi^\prime , z^\prime)$ where $\phi^\prime$ is measured from the $x^\prime$-axis. These are related through the GRT as follows
\begin{eqnarray}\label{rot1}
t^\prime = t \;\;\; , \;\;\; r^\prime = r \;\;\; , \;\;\; \phi^\prime = \phi - \Omega t\;\;\; , \;\;\; z^\prime = z
\end{eqnarray} 
or in its differential form
\begin{eqnarray}\label{rota}
dt^\prime = dt \;\;\; , \;\;\; d r^\prime = d r \;\;\; , \;\;\; d\phi^\prime = d\phi - \Omega dt\;\;\; , \;\;\; dz^\prime = dz
\end{eqnarray} 
It is noted that in both the rotating and non-rotating frames the radial distances are measured from the rotation axis. Through the above equation we would like to emphasize on the meaning of the GRT. Interpreted  {\it kinematically}, as in the cases of linear Galilean and Lorentz transformations,  it introduces a prescription of how the spacetime coordinates of {\it an event} in the two frames are related to one another. This interpretation leads to the following relation between the angular velocities of a test particle observed in the two frames (Fig. 1)
\begin{eqnarray}\label{rotb}
\omega^\prime =  \omega - \Omega
\end{eqnarray} 
which in turn leads to the well-known relation $E^\prime = E - {\bf L} . {\bf \Omega}$ between the energies of the particle in the two frames \cite{Landau0}.
Usually the problem of rotation and rotating frames is discussed in the context of uniformly rotating {\it rigid} disks \cite{Ein}, in other words the rotating frame is a frame attached to a uniformly rotating incompressible disk whose constant angular velocity is measured in the non-rotating (inertial) frame. The above coordinate transformation could also be employed for a uniformly rotating disk and its points (at different times) taken as events whose spacetime coordinates are measured  both in the laboratory frame and in the rotating frame attached to the disk. Obviously in this case it is expected that for any point on the disk $\omega^\prime = 0$ and $\omega = \Omega$ (Fig. 2).\\
\begin{figure}
\begin{center}
\includegraphics[angle=0,scale=0.7]{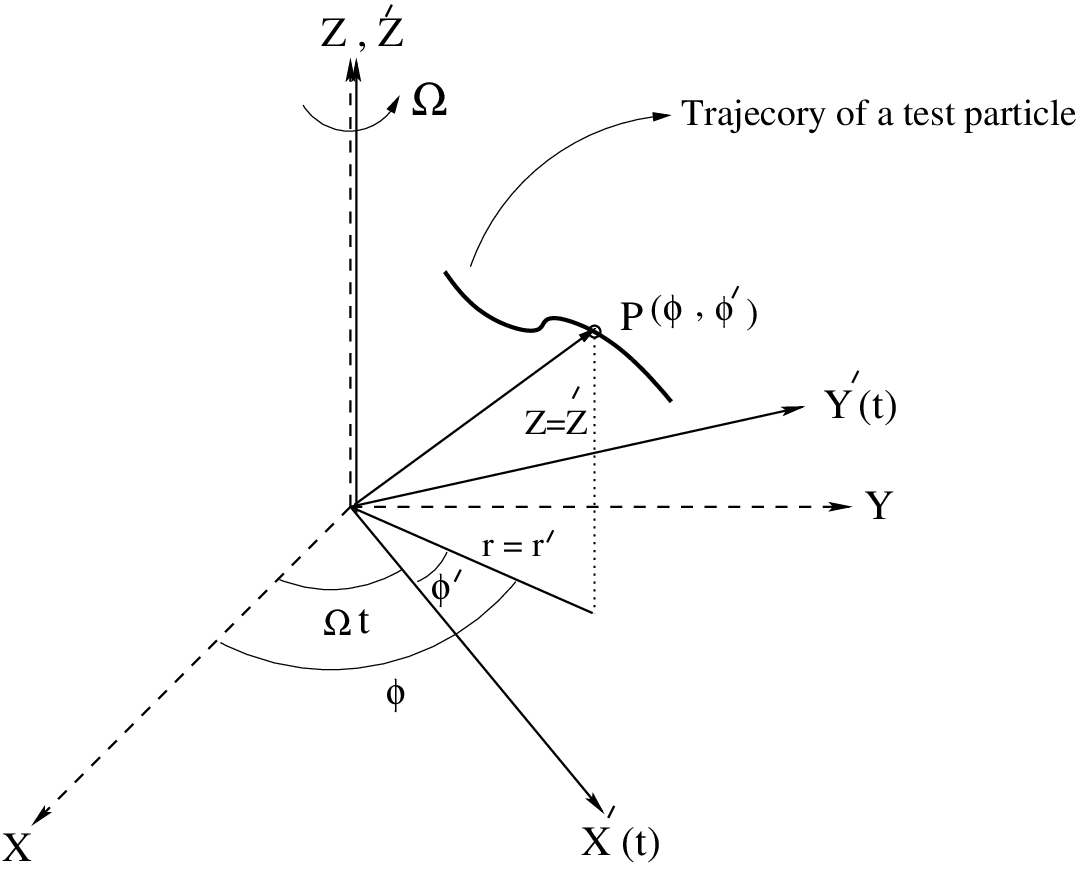}
\caption{Two frames, one rotating (solid) around the other one (dashed) with uniform angular velocity $\Omega$. Trajectory of a test particle and a point $P$ on it as an event observed in the two frames, assigned with angular velocities $\omega$ and $\omega^\prime$.}
\end{center}
\end{figure}
\begin{figure}
\begin{center}
\includegraphics[angle=0,scale=0.7]{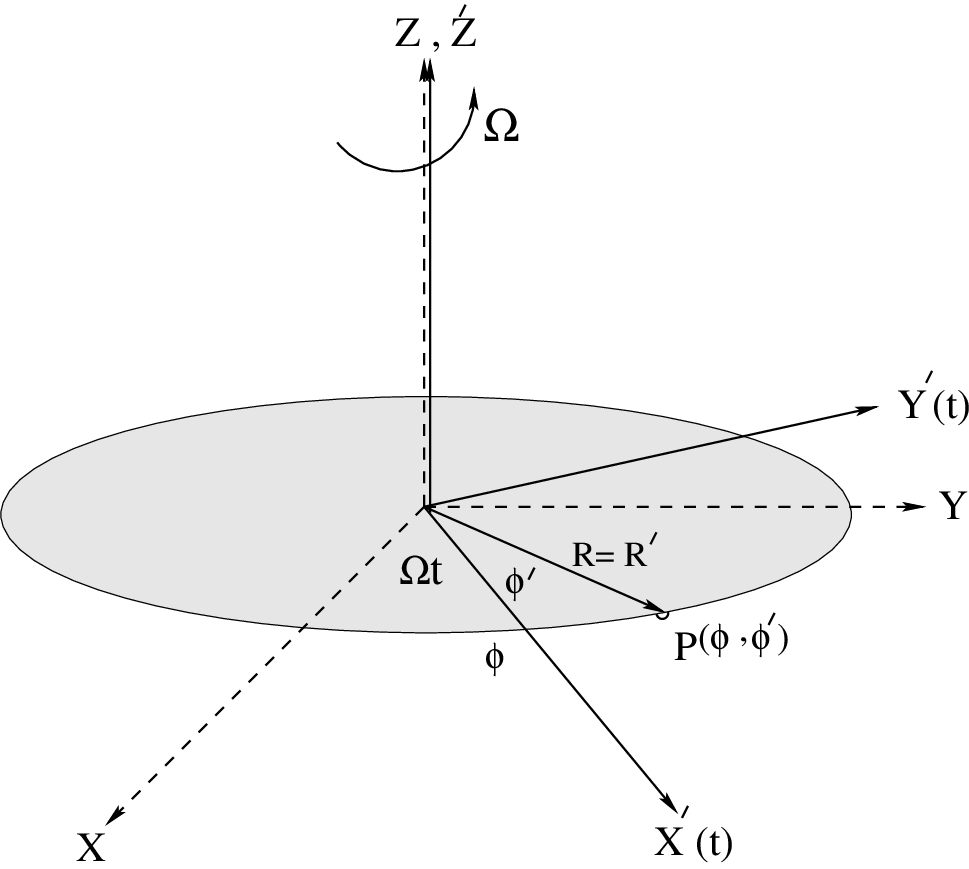}
\caption{A disk and its frame (solid) rotating around the laboratory frame (dashed) with uniform angular velocity $\Omega$. 
Coordinates of a point $P$ in the rim are given in the two frames with angular velocities $\omega^\prime = 0$  and $\omega = \Omega$. }
\end{center}
\end{figure}
\subsection{Ehrenfest's Paradox}
Ehrenfest's Paradox is a contradiction that an {\it inertial (laboratory) observer} faces in applying special relativistic length contraction to a rotating disk. From  an inertial observer's point of view the rim of a rotating disk undergoes a length contraction due to its transverse motion with velocity $v=R\Omega$ and so circumference of a rotating disk ($P^\prime $) is shorter than the one non-rotating ($P$), i.e. $P^\prime < P $. 
On the other hand since the radius of the disk is perpendicular to the direction of the rotational motion of the rim, the same observer will not attribute length contraction to it and so $R^\prime = R$. Therefore the inertial (laboratory) observer, living in a flat spacetime and thereby using the {\it Euclidean} prescription for the circumference of a circle, finds out the contradictory result $P=2\pi R = 2\pi R^\prime = P^\prime$.\\
Perhaps it should be left for experiment to decide which relation holds between $P$ and $P^\prime$ but nevertheless many have tried hard to find either a theoretical resolution to this Paradox or otherwise to invalidate it. An apparently favorite resolution in the literature is based on considering the situation from a rotating observer's point of view and on the idea, introduced by Einstein \cite{Ein,Beren,Gron}, that the spatial geometry in such a frame is {\it non-Euclidean} \footnote{Actually it seems that Theodor Kaluza should be credited with the first assignment of non-Euclidean geometry to a rotating disk \cite{Kalut}, though he has not provided any mathematical detailsto support his idea.}. But, as we will show below, that does not seem to be leading to any kind of resolution of the Paradox  but to a somewhat similar Paradox from the rotating frame's point of view.\\
As pointed out earlier, in the case of a rotating disk one should distinguish between the observer at the center of the disk (called the {\it centrally rotating observer/frame}) whose spatial coordinates, measured in the non-rotating (laboratory) frame, are fixed  and those at different nonzero radii which are  non-inertial due to the centrifugal force felt by them and called {\it orbiting observers/frames}. Einstein calls them {\it eccentric observers} ``relative to whom a gravitational field prevails'' \cite{Ein}.
In other words these observers, by the equivalence principle, find themselves and anything fixed with respect to the disk in a gravitational field. Later, elaborating on this matter, it will be shown that rotating observers at nonzero radii are of central importance in our discussion of RRTs  but for the purpose of Ehrenfest's Paradox we only deal with the rotating observer/frame at the center of the disk. From a rotating observer's point of view  the above-mentioned non-Euclidean character of the disk geometry could be obtained from considering the metric of flat spacetime in the rotating frame, as it is the spatial geometry (metric), defined through spacetime metric, which accounts for spatial distances including that of the disk circumference. Using the  differential GRT (\ref{rota}), the flat spacetime metric in the non-rotating frame
\begin{eqnarray}\label{rot2}
ds^2 = c^2 dt^2 - dr^2 - r^2 d\phi^2 - dz^2
\end{eqnarray} 
transforms into \cite{Landau,Rindler}
\begin{eqnarray}\label{rot3}
ds^2 = (c^2- \Omega^2r^2) dt^2 - 2\Omega r^2 dt d{\phi^\prime} -dr^2 - r^2 d{\phi^\prime}^2 - dz^2
\end{eqnarray} 
in the rotating frame. It is seen that this metric is applicable for radii less than $c/\Omega$, corresponding to the so-called {\it light cylinder}, beyond which $g_{00}$ becomes negative (with the corresponding points having velocities greater than $c$)  and hence from physical point of view  not of interest \cite{Landau,Rindler}.\\ 
The famous result, based on special relativistic arguments made by Einstein, that a rotating clock at  nonzero radius $r = R$ runs slower than that sitting at the center of the disk (or very close to it) \cite{Ein,Beren} is clearly encoded in the above metric, from which we have $d\tau =  \sqrt{1- \frac{\Omega^2 R^2}{c^2}} dt$ where $dt$ is the world time recorded by the inertial/laboratory clocks as well as the one at the center of the disk.
The above spacetime metric plays the same role for a centrally rotating observer that  {\it Rindler spacetime} metric 
\begin{eqnarray}\label{rind}
ds^2 = \eta_{ab} dx^a dx^b = {(1 + a \bar{x}^{1})}^2 {(d\bar{x}^{0})}^2  -{(d\bar{x}^{1})}^2 - {(d\bar{x}^{2})}^2 - {(d\bar{x}^{3})}^2
\end{eqnarray} 
with
\begin{eqnarray}\label{rind1}
x^0 = (a^{-1} + \bar{x}^1)\sinh(a\bar{x}^0)\;\;\;\;\;\;\;\; ; \;\;\;\;\;\;\;\; x^2 = \bar{x}^2 \crcr
x^1 = (a^{-1} + \bar{x}^1)\cosh(a\bar{x}^0)\;\;\;\;\;\;\;\; ; \;\;\;\;\;\;\;\; x^3 = \bar{x}^3
\end{eqnarray} 
plays for a uniformly accelerating observer with 3-acceleration ${\bf a} = (a,0,0)$. In other the words the Rindler metric in the limit $\bar{x}^{1} \ll 1$ (i.e. for points infinitesimally close to the world line of the observer) is equivalent to the Fermi metric \cite{MTW} at first order (i.e. ${\cal O}({\bar x}^l)$) in the absence of rotation (i.e. $\Omega = 0$), while (\ref{rot3}) in the limit $ r \ll 1$ (i.e. infinitesimally close to the centrally rotating observer) is equivalent to the Fermi metric at the same order, in the absence of linear acceleration (i.e. $ a = 0 $) \cite{Nikolic}.
It should be noted that the spacetime in a rotating observer's frame (\ref{rot3}), like Rindler spacetime, is the flat spacetime in a coordinate system which is not maximally extended due to existence of light cylinder in the former and the horizon in the latter. On the other hand, unlike Rindler spacetime, it is a stationary spacetime (reflected in the presence of its cross term $dt d\phi$) and so one needs to employ a spacetime decomposition formalism to define spatial distances and time intervals, and on their basis to prescribe suitable measurement procedures. In what follows we will employ the $1+3$ or {\it threading} formulation of spacetime decomposition \cite {Landau} which is essentially based on sending and receiving light signals between nearby observers (refer to appendix for a brief introduction). Although we are not going to discuss the spacetime measurement procedure here, the employment of the $1+3$ formulation makes it clear that, in principle, we are using light signals to measure the relevant physical quantities, namely spatial distance and time intervals. Based on $1+3$ formulation, the spatial line element for the metric (\ref{rot3}) is given by \cite {Landau} ,
\begin{eqnarray}\label{rot4}
dl^2 = dr^2 + dz^2 + \frac{r^2 d{\phi^\prime}^2}{1-\frac{\Omega^2 r^2}{c^2}}.
\end{eqnarray} 
Now for a circle of radius $r = r^\prime = R$ in the $z=constant$ plane the circumference is given by 
\begin{eqnarray}\label{rot5}
P^\prime = \int _0^{2\pi} dl = \frac{2\pi R}{\sqrt{1-\Omega^2 {R}^2/c^2}} = \frac{P}{\sqrt{1-\Omega^2 {R}^2/c^2}}
\end{eqnarray} 
so that  $P^\prime > P$ with $P$ the circumference of a non-rotating disk. Therefore from the rotating observer's point of view $P$ and $P^\prime$ are also not equal, but the relation between the two quantities is just the opposite of that found by the inertial (laboratory) observer based on Lorentz contraction.\\
The interpretation of the above results goes as follows: Although the transformed spacetime is the flat spacetime in disguise, its spatial geometry now has nonzero Gaussian curvature, leading to the fact that the ratio of the circumference of a circle to its radius is larger than $2 \pi$. We are not going to follow this disagreement on the relation between $P$ and $P^\prime$ from the two observers' points of view nor discuss further the content of Ehrefest's Paradox but there remains a legitimate question that one might ask and that is:
Are we allowed to use the GRT (\ref{rot1}) in all the above considerations? Specially noting that the metric in the rotating frame can be employed out to a specific radius, given by $c/\Omega$ which decreases as we increase the angular velocity. A negative answer to this question has led to the introduction of RRTs.
\subsection{Relativistic rotational transformations}
Our experience with Lorentz transformations intuitively leads to the expectation that GRT is an approximation valid for points near the axis of rotation having small linear (tangential) velocities. Hence for eccentric observers at large radii and/or with high angular velocities one needs to replace the GRT with a relativistic (Lorentz-type) rotational transformation to account for linear (tangential) velocities comparable to $c$. Obviously if one could devise a proper RRT, it might be expected that either the transformation (based on its kinematical interpretation) or the spatial line element of the transformed flat spacetime metric leads to a contracted/dilated circumference for a rotating disk or any other circle of a given radius. \\
A comparison between the usual Lorentz transformation (LT), and GRT is useful at this point. In the case of LT the length contraction is built into the transformation itself and since the flat spacetime line element is form-invariant under the transformation, the length contraction is not expected to be tractable in the form of the corresponding spatial metric. On the other hand in the case of GRT as we noticed, the transformation (\ref{rot1}) is devoid of any length contraction or dilation while the transformed spatial metric (\ref{rot4}) leads to the length dilation. An interesting RRT was introduced by Philip Franklin, a Princeton mathematician, in 1922 \cite{Franklin} and some 30 years later by Trocheris \cite{Trocheris} and Takeno \cite{Takeno} \footnote {In some literature this transformation is called Trocheris-Takeno transformation, but due to Franklin's precedence by almost 30 years and also to highlight his largely overlooked work, we will call it {\it Franklin transformation}.}. Franklin transformation is not the only non-classical rotational transformation and there are a few other proposals such as those introduced in \cite{RRT1}-\cite{RRT5}. RRTs could be classified into two general categories: I- those which employ the same linear velocity distribution as in GRT \cite{RRT2,RRT3,RRT5} and II- those which introduce nonlinear velocity distribution in their construction \cite{Franklin,RRT1,RRT4}. In the former cases the constuction of the RRT is based, in one way or another, on the application of instantaneous Lorentz transformation. For example Post uses the GRT but with a time dilation applying a $\gamma$-factor with a linear velocity distribution \cite{RRT2}, while Strauss modifies Franklin transformation by replacing its nonlinear velocity distribution by a linear one \cite{RRT3}. In \cite{RRT5} the authors introduce an RRT between inertial and non-inertial frames rotating at nonzero radii on circular orbits. Their transformation does not reduce to GRT when the orbit radius is set equal to zero. As an example of the second category, in \cite{RRT1}, Hill introduces an RRT with a nonlinear velocity distribution in terms of Bessel functions which reduces to the classical linear distribution near the rotation axis and approches the upper limit of light velocity at infinity.\\
In the present article we will discuss Franklin transformation and its characteristics including its advantages over the classical transformation and also its drawbacks specially with respect to the corresponding spacetime metric and show how a simple modified version of the transformation could lead to the resolution of some of these drawbacks. Obviously the main criterion for the preference of any non-classical rotational transformation over the classical one (i.e. GRT)  should be the verification of its experimental consequences. For the sake of completeness we will give a brief derivation of Franklin transformation in the next section.
\section{Franklin transformation}
Taking two coordinate frames, $S$ and $S^\prime$, with $S^\prime$ uniformly rotating about $S$, Franklin requires the following plausible conditions and properties to be valid on the relation between the two frames \cite{Franklin}:\\
1-The velocity of a fixed  point in $S^\prime$  with respect to the point in $S$ with which it momentarily coincides is independent of the time, and is the same for all points at a given distance from the axis of rotation.\\
2-For the two concentric circles $r^\prime = r = Constant$, the equations of transformation are similar to those for a Lorentz boost (say along the x-direction) with $r\phi$ the arclength replacing the linear distance (say x).\\
These two properties lead to the  following transformation law 
\begin{eqnarray}\label{rot6}
t^\prime = \gamma(r)\left(t - v(r) r\phi/c^2 \right)\;\;\; ; \;\;\; r^\prime = r\cr
r^\prime \phi^\prime = \gamma(r) \left(r \phi - v(r) t \right)\;\;\; ; \;\;\; z^\prime = z
\end{eqnarray} 
in which $\gamma =\frac{1}{\sqrt{1-v(r)^2/c^2}}$ is the Lorentz-type factor with velocity $v(r)$ to be determined through the last property which is;\\
3-The velocity of a point at the distance $r^\prime + \Delta r^\prime $ from the axis with respect to a point at the distance $r^\prime $ from the axis (both in the system $S^\prime$) is  given by $\Omega \Delta r^\prime $. In other words two different points at two different radii with two different rotational velocities are taken as the analogs of two inertial frames moving uniformly with respect to one another. \\
In effect, the last  property is a prescription for velocity composition law, out of which the nontrivial form of the {\it rotational} velocity is obtained. For two points $B$ and $C$ at radii  $r_B = r $ and $r_C = r + \Delta r $ with velocities $v(r)$ and $v(r + \Delta r)$ (with respect to the point $A$ at the center of the disk) respectively, the composition law reads
\begin{eqnarray}\label{rot6a}
v_{BC} = \frac{v_{AC} - v_{AB}}{1 - \frac{v_{AC}v_{AB}}{c^2}} \Rightarrow \Omega \Delta r = \frac{v(r + \Delta r) - v(r)}{1 - \frac{v(r + \Delta r) v(r)}{c^2}}
\end{eqnarray}
In the limit $\Delta r \rightarrow 0$ this leads to the velocity relation 
\begin{eqnarray}\label{rot7}
v(r) = c \tanh (\Omega r/c)
\end{eqnarray} 
Substituting (\ref{rot7}) in (\ref{rot6}), explicit form of the Franklin transformation (FT) is given by
\begin{eqnarray}\label{rot8}
t^\prime =  \cosh (\Omega r/c)t - \frac{r}{c} \sinh (\Omega r/c) \phi \;\;\; ; \;\;\; r^\prime = r\cr
\phi^\prime = \cosh (\Omega r/c) \phi - \frac{c}{r} \sinh (\Omega r/c) t\;\;\; ; \;\;\; z^\prime = z
\end{eqnarray} 
For points close to the rotation axis i.e. when $\frac{\Omega r}{c} \ll 1$ \footnote{It should  be noted that $\Omega$ is taken as a constant and  such that the integrity of the rotating disk is retained.} this transformation reduces to the classical Galilean transformation by neglecting terms of order $\frac{\Omega^2 r^2}{c^2}$ and higher. These transformations form a group and the inverse transformation is given by changing $\Omega$ to $-\Omega$.
One of the advantages of this transformation over the old Galilean one is in the definition of the velocity given in (\ref{rot7}) which approaches $c$ at
$r\rightarrow \infty$ (i.e. the light cylinder is not at a finite distance but is sent to infinity) and reduces to the Newtonian value $v=\Omega r$ for points near the axis. A formal comparison with a pure Lorentz transformation as a hyperbolic rotation reveals, that it is the linear velocity $v=\Omega r$ in (\ref{rot7}) which now plays the role of some kind of {\it rapidity}.\\
Another obvious difference between Franklin transformation and the Lorentz transformation, when FT is rewritten in the following form,
\begin{eqnarray}\label{rot810}
ct^\prime =  \frac{1}{\sqrt{1-\frac{v(r)^2}{c^2}}}(ct - \frac{v(r)}{c} r\phi) \;\;\; ; \;\;\; r^\prime = r\cr
r \phi^\prime = \frac{1}{\sqrt{1-\frac{v(r)^2}{c^2}}}(r\phi - \frac{v(r)}{c} ct)\;\;\; ; \;\;\; z^\prime = z
\end{eqnarray} 
is the fact that velocity entering the definition of FT unlike LT is not a constant but an $r$-dependent quantity. This will lead to undesirable results in the case of FT when we consider the transformed spacetime metric (i.e. in the rotating frame) and the corresponding spatial distances and time intervals. It will be shown that neither will reduce to their expected  expressions at small rotational velocities (i.e. when $\frac{\Omega r}{c} \ll 1$). But before discussing these issues, it seems appropriate to discuss  interpretation of FT as compared to those of GRT and LT.
\subsection{Interpretation of FT}
An important issue about the Franklin transformation, which seems to be taken for granted in most of the previous studies, is its interpretation as
the transformation of the spacetime coordinates of an {\it event} between two frames; a non-rotating (inertial) frame and another one rotating 
uniformly about their common axis. This is the same usual interpretation attributed to the GRT as illustrated in Fig. 1. 
But characteristics of FT would prevent one to easily interpret this transformation as a kinematical one. The main characteristic acting so is the radial dependence of velocity entering the transformation. This velocity distribution is attributed to the rigid arms of the rotating frame (or disk points if the frame is attached to a uniformly rotating rigid disk) and so, taking into account the fact that in FT the non-rotating and rotating frames share the rotation axis,  the transformation of the arclengths in FT ( which is given in terms of this velocity) is only valid for disk points.
By the above reasoning, it seems more reasonable to look at  FT as  a transformation specially tailored for the problem of a rotating disk in which events are nothing but different points of a rotating disk at different times. In other words one should be cautious in interpreting FT as a kinematical transformation relating coordinates of an event in a rotating frame to  that of an inertial non-rotating one. For example, based on  kinematical interpretation of FT, for events on the rotation axis i.e. for  $r = r^\prime = 0$ (where the cylindrical coordinate system is degenerate and $v(0)= 0$), FT reduces {\it exactly} to GRT and this has no clear interpretation. If FT is going to be elevated to a kinematical transformation one needs to modify and reinterpret it.
\section{spacetime Metric and spatial geometry in the rotating frame through a Franklin transformation}
Using the inverse of the Franklin transformation in its differential form
\begin{eqnarray}\label{rot8a}
c dt =  \cosh (\Omega r/c) c dt^\prime + r \sinh (\Omega r/c) d\phi^\prime + A_1 dr\;\;\; ; \;\;\; dr = dr^\prime \crcr
r d\phi =  \cosh (\Omega r/c) r d\phi^\prime + \sinh (\Omega r/c) c dt^\prime + A_2 dr\;\;\; ; \;\;\; dz = dz^\prime \crcr
A_1 = \sinh (\Omega r/c) (\phi^\prime + \Omega t^\prime) + \cosh (\Omega r/c)(\frac{\Omega r}{c} \phi^\prime)\crcr
A_2 = \sinh (\Omega r/c) (\frac{\Omega r}{c} \phi^\prime - c t^\prime/r) + \cosh (\Omega r/c)(\Omega t^\prime)
\end{eqnarray}
and substituting in (\ref{rot2}) the spacetime metric in the rotating frame is given by
\begin{eqnarray}\label{rot8b}
ds^2 = c^2 {dt^\prime} ^2 - (1 - A_1^2 + A_2^2 ) dr^2 - r^2 d{\phi^\prime}^2 - dz^2  
+ \crcr 2 c \left( A_1 \cosh (\Omega r/c)- A_2 \sinh (\Omega r/c)\right) dr dt^\prime 
+ 2 \left( A_1\sinh (\Omega r/c) - A_2 \cosh (\Omega r/c)\right) r dr d{\phi^\prime}
\end{eqnarray}
Unlike the cross term in (\ref{rot3}) which is the typical $dt^\prime d{\phi^\prime}$ term representing the rotational character of the metric, the cross terms in the above metric include $dr dt^\prime$ and  $dr d{\phi^\prime}$ terms and that is why the reduction of this metric form 
to (\ref{rot3}) for $\Omega r/c \ll 1$ is not expected. Also it should be noted that due to the explicit appearance of $\phi^\prime$ 
and $t^\prime$ in (\ref{rot8b}) both the {\it temporal} and {\it angular} isometries present in (\ref{rot3}) are now lost.
\subsection{Spatial distances and time intervals}
From the above result on the spacetime metric it is obviously not expected that the spatial geometry corresponding to (\ref{rot8b}) is reducible to 
the one given by (\ref{rot4}) in the limit $\Omega r/c \ll 1$. Indeed using the $1+3$ decomposition (Eq. (\ref{appa2})) the spatial metric corresponding to (\ref{rot8a}) is given by
\begin{eqnarray}\label{rot8c}
dl^2 = \{1 - A_1^2 + A_2^2  + 4c^2 [ A_1 \cosh (\Omega r/c)- A_2 \sinh (\Omega r/c)]^2\} dr^2 + \crcr
-2 [ A_1\sinh (\Omega r/c) - A_2 \cosh (\Omega r/c)] r dr d{\phi^\prime} + dz^2 + {r^2 d{\phi^\prime}^2}
\end{eqnarray}
through which the circumference of a disk with radius $r=R$ in the $z=constant$ plane is given by the Euclidean value $2 \pi R$ compared to the non-Euclidean value (\ref{rot5}) obtained through the Galilean transformed spatial metric (\ref{rot4}). It should be noted that despite the above fact the Gaussian curvature of the spatial metric is not zero indicating the non-Euclidean nature of the spatial metric \cite{Franklin}.
It should also be noted from (\ref{rot8b}) that proper time interval in the rotating frame, for a clock fixed at $r=R$, is given by  
\begin{eqnarray}\label{rot81c}
 d\tau^\prime =  dt^{\prime} =  \cosh^{-1} (\Omega R/c) dt
\end{eqnarray}
where use is made of (\ref{rot8a}). In the limit  $\frac{\Omega R}{c} \ll 1$ the above relation reduces to that obtained from the Galilean transformed metric for rotating clocks at nonzero radii i.e. $d\tau^\prime = \sqrt{1-\frac{\Omega^2 R^2}{c^2}} dt$. On the other hand, as we discussed earlier, one could relate spatial distances and time intervals  not only through the metric obtained from Franklin transformation but also through the coordinate transformations themselves according to their kinematical interpretation. Obviously using the formal analogy between FT and LT one can obtain relation between spatial distances (arclengths) and time intervals in the two coordinate systems as follows:
\begin{eqnarray}\label{rot81}
\Delta t = \frac{1}{\sqrt{1- \frac{v^2}{c^2}}} \Delta t^\prime
\end{eqnarray}
\begin{eqnarray}\label{rot82}
\Delta l = R\Delta \phi = \sqrt{1- \frac{v^2}{c^2}} R\Delta\phi^\prime =  \sqrt{1- \frac{v^2}{c^2}} \Delta l^\prime,
\end{eqnarray}
where in (\ref{rot81}) we employed $\Delta \phi^\prime = 0$ (see Fig. 2) and in (\ref{rot82}) used the simultaneous measurements ($\Delta t = 0$) of both ends of the corresponding arclength. The above equations correspond to the time dilation and length contraction of clocks and rulers at rest in the rotating observer's frame $S^\prime$ respectively. With $v=c\tanh (\frac{\Omega R}{c})$ at radius $r=R$, the above results are consistent with what one expects from applying special relativistic length contraction (based on LT) to a rotating disk for $\frac{\Omega R}{c} \ll 1$. It seems that once again we are faced with the Ehrenfest's Paradox, in the sense that using the spatial geometry given by Eq. (\ref{rot8c}) implies that the circumference of a rotating disk is the same as the circumference of the non-rotating disk whereas, employing Franklin transformation, the circumference of a rotating disk is found to be shorter than the one non-rotating. 
\subsection{Angular velocity of a test particle/disk point in the two frames related by FT}
Using the differential Franklin transformation (\ref{rot8a}) to calculate  the rotational frequency in the inertial observer's frame we find
\begin{eqnarray}\label{rot8d}
\omega = \frac{d\phi}{dt} = \frac{ \cosh (\Omega r/c) d\phi^\prime + \frac{c dt^\prime}{r}\sinh (\Omega r/c)  + \frac{A_2}{r}dr^\prime} { \cosh (\Omega r/c) dt^\prime + \frac{r}{c}\sinh (\Omega r/c) d\phi^\prime + \frac{A_1}{c}dr^\prime} 
\end{eqnarray}
from which for the frequency in the rotating frame we have
\begin{eqnarray}\label{rot8k}
\omega ^\prime =  \frac{d\phi^\prime}{dt^\prime} = \frac{\omega \cosh(\Omega r/c) - \frac{c}{r} \sinh(\Omega r/c) + \frac{dr}{dt^\prime}(\frac{A_1}{c} \omega - \frac{A_2}{r})}{\cosh(\Omega r/c) - \omega \frac{r}{c} \sinh(\Omega r/c)}
\end{eqnarray}
In the limit where $(\Omega r/c) \ll 1$, the above expression reduces to the classical relation (\ref{rotb})
\begin{eqnarray}\label{rot8f}
\omega ^\prime \approx \omega - \Omega.
\end{eqnarray}
\section{Modified Franklin transformation: its interpretation and the spacetime metric in the rotating frame}
As it is obvious from its derivation, Franklin transformation was obtained in close analogy with the usual Lorentz transformation for inertial frames moving with {\it constant velocities} relative to one another. Our starting point for modification of Franklin transformation is its main formal difference from the Lorentz transformation which is the dependence of relative velocity on the radial coordinate (i.e. $v \equiv v(r)$) in (\ref{rot7})). It is clear from Franklin's derivation of (\ref{rot7}) that this coordinate-dependent velocity is a direct consequence of applying the relativistic composition law to  high rotational velocities. Indeed the nonlinear velocity relation (\ref{rot7}) could also be obtained by the requirement that for any two infinitesimally close points (separated by a radial distance $dr$) on a uniformly rotating rigid rod (divided into $n$ infinitesimal segments), the difference in  their linear velocities is given by $\Omega d r$ \cite{Cao}. Then using the relativistic composition law iteratively to find the velocity at a finite distance along the rod,  in the limit $n\rightarrow \infty$, one ends up with the velocity distribution (\ref{rot7}).
Since the kinematical transformation is supposed to give the relation between coordinates assigned to events by two observers, an inertial non-rotating one (laboratory observer/frame) and a non-inertial rotating observer at a given radius $R$, going back to the transformation law (by formal analogy with LT), it is the observer velocity at that radius (i.e. $v = c \tanh(R\Omega/c)$) which should enter the transformation law. Indeed it has already been pointed out in some literature \cite{Kichen1,Kichen}, without further clarification, that Franklin transformation leads to inconsistencies if one neglects the fact that it is determined at $r=constant$ as well as at $z=constant$. We have mentioned some of these inconsistencies  in the previous sections, and so by the above argument we introduce the following modified Franklin transformation (MFT)
\begin{eqnarray}\label{rot9}
t^\prime =  \cosh (\Omega R/c)t - \frac{R}{c}  \sinh (\Omega R/c)\phi \;\;\; ; \;\;\; r^\prime = r \cr
\phi^\prime =  \cosh (\Omega R/c) \phi - \frac{c}{R} \sinh (\Omega R/c) t\;\;\; ; \;\;\; z^\prime = z 
\end{eqnarray}
This could be obtained by changing the second and third steps in the derivation of the Frankiln transformation  by assigning observers to the disk points at a given radius $r=R$, for which the velocity with respect to the inertial observers, using the third step, is found to be $v = c \tanh(R\Omega/c)$. In terms of this velocity the MFT could be written as follows
\begin{eqnarray}\label{rot811}
ct^\prime =  \frac{1}{\sqrt{1-\frac{v^2}{c^2}}}(ct - \frac{v}{c} R\phi) \;\;\; ; \;\;\; r^\prime = r\cr
R \phi^\prime = \frac{1}{\sqrt{1-\frac{v^2}{c^2}}}(R\phi - \frac{v}{c} ct)\;\;\; ; \;\;\; z^\prime = z
\end{eqnarray} 
This is indeed a simple, physical modification with profound consequences. To see its effects, first of all we find the equivalent metric by finding the inverse differential transformation which is 
\begin{eqnarray}\label{rot10}
dt =  \cosh (\Omega R/c)dt^\prime + \frac{R}{c}  \sinh (\Omega R/c)d\phi^\prime \;\;\; ; \;\;\; dr = dr^\prime \cr
d\phi =  \cosh (\Omega R/c) d\phi^\prime + \frac{c}{R} \sinh (\Omega R/c) dt^\prime\;\;\; ; \;\;\; dz = dz^\prime 
\end{eqnarray}
and substituting them in the inertial frame's flat spacetime metric (\ref{rot2}) upon which we end up with 
(taking $\beta = \frac{R \Omega}{c}$)
\begin{eqnarray}\label{rot11}
ds^2 = c^2\cosh^2\beta(1- \frac{{r}^2}{R^2}\tanh^2\beta) {dt^\prime}^2 - {dr}^2 -  
{r}^2 \cosh^2\beta  \cr (1- \frac{R^2}{{r}^2}\tanh^2\beta)d{\phi^\prime}^2   
+ 2cR \sinh \beta \cosh \beta (1- \frac{{r}^2}{R^2})dt^\prime d\phi^\prime  - {dz}^2
\end{eqnarray} 
Note that now there is a radial coordinate $r$ as well as a constant radius $R$  which specifies a class of observers fixed at that radius. This will allow a kinematical interpretation of the above MFT. In other words no matter what the constant radius in (\ref{rot9}), this transformation gives a prescription of how the temporal ($t \; \& \; t^\prime$) and angular ($\phi \; \& \; \phi^\prime$) coordinates of an event in the two frames are related. Indeed, it is now that one could justify the division of the originally introduced transformation of {\it arclengths} ($r^\prime \phi^\prime \; \& \; r\phi$ for an event at radial coordinate $r=r^\prime$) by the common radial coordinate leading to the transformation of {\it angular coordinates} $\phi \; {\rm and} \; \phi^\prime$. In other words the angular coordinates are defined using the arclengths at the radial position $r=r^\prime= R$ of the eccentric observer. It should be noted that spatial coordinate measurements by the inertial as well as the eccentric (non-inertial) observers are made from the axis of rotation as a {\it preferred direction} and the eccentric observers carry their own clocks but use the triad axes of the centrally rotating observer to designate spatial coordinates to events.
The presence of $R$ as a constant in the transformed flat spacetime as given by (\ref{rot10}) may look strange but obviously it is no stranger than the appearance of $\Omega$ in (\ref{rot1}) or in (\ref{rot8}). Both $\Omega$ and $R$ are transformation parameters, one ($\Omega$) from an inertial observer's frame to a centrally rotating frame and the other ($R$) from the centrally rotating observer's frame to a set of equivalent rotating observers at radius $R$ (non-inertial observers). Indeed they are  now combined to form the new transformation parameter which is the velocity $v = c \tanh(R\Omega/c)$ (or $R\Omega$ for that matter). Also compared to the case of Rindler metric, in which the observer's acceleration  enters the spacetime metric (\ref{rind}), the appearance of the parameter $R$ which determines an eccentric observer's velocity and acceleration is expected on the same grounds. Further it should not be forgotten that the {\it spacetime} in the rotating coordinates is always flat, for a coordinate transformation never changes the nature of a spacetime whether it is the old Galilean transformation (\ref{rot3}) or FT (both having the parameter $\Omega$) or MFT (with parameter $R\Omega$), and it is only the {\it spatial metric} in the rotating observer's frame which loses its Euclidean character.
Obviously the metric  (\ref{rot11}) is of interest for radial distances 
\begin{eqnarray}\label{rot11-0}
r \leqslant \frac{\beta}{|\tanh \beta|}(\frac{c}{\Omega}),
\end{eqnarray} 
and in the classical Galilean limit where $\beta \ll 1$ (i.e. close to the rotation axis) it reduces to
\begin{eqnarray}\label{rot11-1}
ds^2 = c^2 (1- \frac{r^2 \Omega^2}{c^2}) {dt^\prime}^2 - {dr}^2 -  
r^2 (1- \frac{R^2}{r^2}\beta^2)d{\phi^\prime}^2 
+ 2R^2 \Omega (1- \frac{r^2}{R^2})dt^\prime d\phi^\prime  - {dz}^2,
\end{eqnarray} 
which in turn reduces to the spacetime metric (\ref{rot3}) under the extra condition that the radial coordinates of the events under consideration are larger than or equal to $R$. In other words, for observers close to the axis the range  $R \leqslant r < \frac{c}{\Omega}$ replaces the range $0 \leqslant r < \frac{c}{\Omega}$ \footnote{Note that the condition $\beta \ll 1$ is equivalent to $R \ll \frac{c}{\Omega}$ whereas the same condition employed in (\ref{rot11-0}) leads to $r \leqslant \frac{c}{\Omega}$.}. So, unlike the Franklin transformation, not only the transformation itself, but also the metric in rotating frame reduces to the Galilean one in the limit $\beta \ll 1$. 
It should be noted that for $r = R$ in (\ref{rot11}), i.e.  at the radial position of the eccentric observer, the metric reduces to that of a spatially Euclidean flat spacetime (\ref{rot3}) of an inertial observer, i.e. at $r = R$ the form of the spacetime metric is invariant under MFT. This is a feature of (\ref{rot11-1}) which is somewhat shared with the Fermi metric of an accelerated, spinning observer in  flat or curved background. Recall the feature of the Fermi metric that on the observer's world line reduces to the Minkowski metric \cite{MTW}.
Now the reduction of MFT to exact GRT, by setting $R=0$ in (\ref{rot9}), while (\ref{rot11}) reduces to (\ref{rot3}), has a consistent interpretation (in contrast to setting $r=0$ in FT which was shown to lead to inconsistencies with respect to its kinematical interpretation); it corresponds to the centrally rotating observer who is at rest with respect to the non-rotating inertial (laboratory) observer, and so their observations are naturally related through GRT. So in our setting of the problem of rotation and rotating frames, we have drastically changed the scenario by introducing non-inertial observers fixed at nonzero radii on the disk and also introducing the MFT as the kinematical transformation between the coordinates assigned to events by these observers and the inertial ones.\\
In the next two subsections we find how the spatial and time intervals in the rotating and inertial frames are related through MFT. We also discuss energy and angular velocity of a test particle (disk point) in the two frames. It should be noted that the eccentric observers use a local Cartesian coordinate system attached to a rigidly rotating disk at their position such that its axes are always parallel to the axes of the Cartesian coordinte system used by the centrally rotating observer. In this way the radial coordiantes assigned to events by all observers are  measured from the rotation axis. 
\subsection{Spatial line element and spatial distances}
Using the $1+3$ approach (Appendix A), the metric  (\ref{rot11}) could be written in the following form
\begin{eqnarray}\label{rot11a}
ds^2 = c^2\cosh^2\beta(1- \frac{r^2}{R^2}\tanh^2\beta) \left({dt^\prime} - A_\alpha d{x^\prime}^\alpha\right)^2 - dl^2 
\end{eqnarray} 
in which the spatial line element is given by
\begin{eqnarray}\label{rot11b}
dl^2 = dr^2 + dz^2 + \left( r^2\cosh^2\beta(1-\frac{R^2}{r^2}\tanh^2\beta) + 
R^2 \frac{\sinh^2\beta(1-\frac{r^2}{R^2})^2}{(1-\frac{r^2}{R^2}\tanh^2\beta)}\right){d\phi^\prime}^2
\end{eqnarray} 
and the gravitomagnetic potential is
\begin{eqnarray}\label{rot11c}
A_\alpha \equiv A_{\phi^\prime} \delta^{\phi^\prime}_\alpha = (0, 0, -R \frac{\tanh\beta(1-\frac{r^2}{R^2})}{(1-\frac{r^2}{R^2}\tanh^2\beta)})
\end{eqnarray} 
Now one could find the circumference of a circle/disk of radius $r$ in $z= constant$ plane using the above line element as
\begin{eqnarray}\label{rot11d}
L_{MFT} = \int dl = \int_0^ {2 \pi} \left( r^2\cosh^2\beta(1-\frac{R^2}{r^2}\tanh^2\beta) + 
R^2 \frac{\sinh^2\beta(1-\frac{r^2}{R^2})^2}{(1-\frac{r^2}{R^2}\tanh^2\beta)}\right)^{1/2}{d\phi^\prime}
\end{eqnarray} 
It is an easy task to show that the above spatial line element (\ref{rot11})  reduces to the classical spatial element (\ref{rot4}) in the limit of $\beta \ll 1$ . Also it is noted that for an observer fixed at nonzero radius $R$, a circle at that radius i.e. $r=R$, has the Euclidean circumference $2\pi R$ as expected from the form invariance of the metric (\ref{rot11}) at that radius. 
On the other hand using the  MFT (\ref{rot10}), one obtains the following relation between the differential arclengths (at radius $R$) as measured by the rotating and inertial observers:
\begin{eqnarray}\label{rot12d}
R d\phi =  \cosh (\Omega R/c) R d\phi^\prime
\end{eqnarray}
In other words as in the case of FT, again we are faced with the Ehrenfest's Paradox in the sense that an arclength of a rotating disk, measured by the inertial observer, is the same as that of the non-rotating disk if spacetime metric is employed but different if MFT is used. The {\it relation} between length measurements by the inertial and rotating observers, based on MFT and hypothesis of locality \cite{Mash2}, are discussed and compared in \cite{Nouri1}.
\subsection{Time intervals and their relations}
As in the case of FT one can obtain the relation between proper time intervals in the inertial observer's frame and that of a rotating one at a nonzero radius $R$ using the MFT. Using the MFT or its corresponding metric (\ref{rot11}) in the rotating frame, we find that the proper time intervals at 
rest frame of the clock at $r=R$  and that at the center of the disk $r = 0$ (measured by an inertial observer)  are related by,
\begin{eqnarray}\label{13e}
{d\tau_0}= \cosh\beta {d\tau},
\end{eqnarray}
corresponding to time dilation of a rotating clock readings as measured by an observer in the inertial frame. This is the same relation obtained in the case of FT (see Eq. (\ref{rot81c})). In the limit where $\beta \ll 1$ the above relation for finite time intervals reduces to 
\begin{eqnarray}\label{13f}
{\Delta \tau_0} \approx (1 + \frac{\Omega^2}{2c^2}{R}^{2} +
\frac{5}{24}\frac{\Omega^4}{c^4}{R}^{4}) {\Delta \tau},
\end{eqnarray}
which up to the second order in $\beta$ agrees with the relation based on applying instantaneous Lorentz transformation along with linear velocity distribution employed in GRT \cite{Nouri1}. An application of the instantaneous Lorentz transformation is experimentally verified in the measurements of circulating muons lifetime at CERN \cite{Bailey}, but one should be cautious that in applying MFT the ticking clock is fixed at a nonzero radius on a {\it rotating platform} and not forced to move on a circular path by the application of electromagnetic fields. So if one is going to test the above theoretical prediction in an experimental setup, it should be a setup with an unstable particle fixed at a nonzero radius on a rotating platform. The same argument as above could be used to discuss the transverse Doppler effect as a rotational phenomenon in the context of MFT \cite{Nouri1}.
\subsection{Energy of a test particle}
The energy of a  particle of mass $m$ moving with 3-velocity $v$ in a stationary field  was shown in the Appendix A to be given by,
\begin{eqnarray}\label{13g}
E  = \frac{mc^2 \sqrt{g_{00}}}{\sqrt{1-\frac{v^2}{c^2}}},
\end{eqnarray}
which is a conserved quantity. For a particle fixed at a constant radius $R$ on the rotating frame (e.g. on a rigidly rotating disk), in the comoving frame (i.e. $v=0$) which is the rotating frame of the eccentric observer at $R$, the same energy is given by $E^\prime =mc^2 \sqrt{g_{00}(r=R)}$ so that Eq. (\ref{13g}) could be rewritten as follows:
\begin{eqnarray}\label{13h}
E = \frac{E^\prime}{\sqrt{1-\frac{v^2}{c^2}}}.
\end{eqnarray}
Now using the fact that in MFT the 3-velocity at radius $R$ is given by $v=c\tanh(R\Omega/c)$,  the above relation reduces to,
\begin{eqnarray}\label{13k}
E = \cosh(R\Omega/c) {E^\prime} \approx (1 + \frac{\Omega^2}{2c^2}{R}^{2} +
\frac{5}{24}\frac{\Omega^4}{c^4}{R}^{4}) {E^\prime}
\end{eqnarray}
which is again, up to the second order in $\beta$, in  agreement with the relation based on applying instantaneous Lorentz transformation along with the linear velocity distribution on a uniformly rotating disk. Obviously both the above result and the relation (\ref{13f}), are direct consequences of the nonlinear velocity distribution $v = c\tanh (R\Omega/c)$ on the disk.
\subsection{Angular velocity of a test particle/disk in MFT}
In terms of the kinematical interpretation, the relation betweeen the angular velocities of a test particle in the two frames, related by the MFT, is found by employing the inverse differential rotation (\ref{rot10}) so that 
\begin{eqnarray}\label{rot11f}
\omega = \frac{d\phi}{dt} = \frac{ \cosh \beta d\phi^\prime + \frac{c}{R}\sinh \beta dt^\prime} {\cosh \beta dt^\prime + \frac{R}{c}\sinh \beta d\phi^\prime} 
\end{eqnarray}
leading to 
\begin{eqnarray}\label{rot11f1}
\omega^\prime = \omega (1 + \frac{R}{c}\tanh \beta) - \frac{c}{R}\tanh \beta
\end{eqnarray}
in which we used the fact that $\omega^\prime = \frac{d\phi^\prime}{dt^\prime}$. As in  the case of FT, it could easily be seen that in the limit of $\beta \ll 1$ the above relation reduces to  the classical relation (\ref{rotb}) which was found through the Galilean transformation.
On the other hand, from an inertial observer's point of view, angular velocity of the disk (or its points) is $\frac{d\phi}{dt}=\Omega$ and so the above relation, for the disk itself changes into
\begin{eqnarray}\label{rot11g}
\omega ^\prime = \Omega (1 + \frac{R}{c}\tanh \beta) - \frac{c}{R}\tanh \beta
\end{eqnarray}
in other words, in MFT, for the eccentric observers, the angular velocity of the rotating disk depends on the radial position of the observer. But close to the centrally rotating observers, i.e. in the limit  $\beta \ll 1$, the expectation based on GRT is restored  where $\omega ^\prime \approx 0$.
\section{Non-invariance of electromagnetism under (modified) Franklin transformation}
In some of the studies in the literature discussing the Franklin transformation it is claimed that this transformation restores the full Lorentz (-type) covariance of electrodynamics \cite{Kichen,Hillion}. Here we show in detail that such a claim is not correct and 
the covariance  mentioned in those studies only is satisfied by implicitly fixing the radial coordinate in the transformation (i.e. $r=constant$), in which case the transformed metric \eqref{rot8b} retains its Euclidean form by setting $dr=0$.  But for a general transformation this is not true as shown below for the MFT, in which case again, the covariance is  only restored at the position of the observer i.e. at $r=R$ where the spacetime, as discussed and interpreted previously, is Euclidean form invariant.\\
To be specific, under Lorentz transformation, Maxwell equations are invariant in the sense that they retain the same three-dimensional vector form in the transformed coordinates, consequently the electromagnetic wave equation which is obtained from these equations is also form invariant. In what follows we show that neither the Maxwell equations nor wave equation are form invariant under Franklin transformation. To make life easier we show this in the absence of any EM sources and for the modified Franklin transformation, but the same result (non-invariance of electromagnetism) holds for the original Franklin transformation.
From modified Franklin transformation (\ref{rot9}) we have the following relation between the partial derivatives:
\begin{eqnarray}\label{rot12}
\frac{\partial}{\partial t^\prime } = \cosh\beta \frac{\partial}{\partial t} + \frac{1}{R} \sinh\beta \frac{\partial}{\partial \phi }\cr \frac{\partial}{\partial \phi^\prime } = R\sinh\beta \frac{\partial}{\partial t}+ \cosh\beta \frac{\partial}{\partial \phi }\cr \frac{\partial}{\partial r^\prime } =  \frac{\partial}{\partial r}\;\;\;\; ; \;\;\;\; \frac{\partial}{\partial z^\prime } =  \frac{\partial}{\partial z}  
\end{eqnarray}
\subsection{Non-invariance of wave equation under MFT }
Using the above relations the wave equation in the unprimed coordinates (inertial frame)
\begin{eqnarray}\label{rot13}
\frac{\partial^{2}\psi}{\partial{{t}^{2}}}-\frac{1}{r}\partial_{r}(r{\frac{\partial\psi}{\partial r}})-
\frac{1}{r^{2}}\frac{\partial^{2}\psi}{\partial{{\phi}^{2}}}-\frac{\partial^{2}\psi}{\partial{z^{2}}}=0\
\end{eqnarray}
transforms into
\begin{eqnarray}\label{rot14a}
(\frac{r^{2}\cosh^{2}\beta-R^2\sinh^{2}\beta}{r^2})\frac{\partial^{2}\psi}{\partial{{t^\prime}^{2}}}+
2(\frac{(R^2-r^{2})\sinh\beta\cosh\beta}{Rr^2})\frac{\partial^{2}\psi}{\partial{t^\prime}\partial{\phi^\prime}}
-\frac{1}{r}\partial_{r}(r{\frac{\partial\psi}{\partial r}})\nonumber\\+
(\frac{r^{2}\sinh^{2}\beta-R^2\cosh^{2}\beta}{R^2r^2})\frac{\partial^{2}\psi}{\partial{{\phi^\prime}^{2}}}-\frac{\partial^{2}\psi}{\partial{z^{2}}}=0
\end{eqnarray}
under MFT, i.e. the wave equation is not form invariant under MFT. The same result could also be obtained by using the metric corresponding to MFT (Eq. (\ref{rot11})) and the following general form of the wave equation in a curved background with metric $g_{ij}$
\begin{equation}\label{rot15a}
{\Box}{\psi}=\frac{1}{\sqrt{g}}\frac{\partial}{\partial{q_i}}(g^{1/2}g^{ik}\frac{\partial\psi}{\partial{q_k}})=0
\end{equation}
where $q_i=t^\prime,r,\phi^\prime,z$.
\subsection{Non-invariance of Maxwell equations under MFT }
To obtain (source-free) Maxwell equations for a rotating observer from those in the frame of an inertial observer related through MFT
we use the field tensor in the spacetime of a rotating observer (MFT metric) given by:
\begin{eqnarray}\label{rot16a}
{F^\prime}_{ij}=\left(
                  \begin{array}{cccc}
                    0 & -\frac{A}{R}{E^\prime}_r & -{r}{E^\prime}_{\phi^\prime} & -\frac{A}{R}{E^\prime}_z \\
                    \frac{A}{R}{E^\prime}_r & 0 & -\frac{\tilde {A}}{A}{E^\prime}_r+\frac{R{r}}{A}{B^\prime}_z & -{B^\prime}_{\phi^\prime} \\
                    {r}{E^\prime}_{\phi^\prime} & \frac{\tilde {A}}{A}{E^\prime}_r-\frac{R{r}}{A}{B^\prime}_z & 0 & \frac{\tilde {A}}{A}{E^\prime}_z+\frac{R{r}}{A}{B^\prime}_r \\
                    \frac{A}{R}{E^\prime}_z & {B^\prime}_\phi & -\frac{\tilde {A}}{A}{E^\prime}_z-\frac{R{r}}{A}{B^\prime}_r & 0 \\
                  \end{array}
                \right)
\end{eqnarray}
where
\begin{equation}\label{rot17}
A=\sqrt{R^2\cosh^{2}\beta-r^{2}\sinh^{2}\beta}~~~~{\rm and}~~~~\tilde {A}=(-R^2+r^{2})\sinh\beta\cosh\beta
\end{equation}
so that the inhomogeneous equations
\begin{equation}\label{rot18}
\frac{1}{\sqrt{g}}~\partial_{i}(\sqrt{g}{F^\prime}^{ij})= 0
\end{equation}
are given by
\begin{eqnarray}\label{rot19}
\partial_r[r(\frac{R}{A}{E^\prime}_r-\frac{\tilde {A}}{r{A}}{B^\prime}_z)]+\partial_{\phi^\prime}({E^\prime}_{\phi^\prime})+\partial_z[r(\frac{R}{A}{E^\prime}_z+\frac{\tilde {A}}{A}{B^\prime}_r)]=0\\
\frac{R}{A}\partial_{t^\prime} {E^\prime}_r-\frac{\tilde {A}}{r{A}}\partial_{t^\prime} {B^\prime}_z-\frac{A}{r{R}}\partial_{\phi^\prime}{{B^\prime}_z}+\partial_z{{B^\prime}_{\phi^\prime}}=0\\
\partial_{t^\prime} {E^\prime}_{\phi^\prime}+\partial_r(\frac{A}{R}{B^\prime}_z)-\partial_z(\frac{A}{R}{B^\prime}_r)=0\\
\frac{r{R}}{A}\partial_{t^\prime} {E^\prime}_z+\frac{\tilde {A}}{A}\partial_{t^\prime} {B^\prime}_r-\partial_r(r{{B^\prime}_{\phi^\prime}})
+\frac{A}{R}\partial_{\phi^\prime}{{B^\prime}_r}=0
\end{eqnarray}
respectively for $j=0,1,2,3$. Also the homogeneous equations
\begin{equation}
\partial_{[i}{{F^\prime}_{jk]}}=0
\end{equation}
give rise to 
\begin{eqnarray}\label{rot20}
\partial_r(\frac{\tilde {A}}{A}{E^\prime}_z+\frac{R{r}}{A}{B^\prime}_r)+\partial_{\phi^\prime}{{B^\prime}_{\phi^\prime}}+\partial_z(-\frac{\tilde {A}}{A}{E^\prime}_r+\frac{R{r}}{A}{B^\prime}_z)=0\\
\partial_{t^\prime}(-\frac{\tilde {A}}{A}{E^\prime}_r+\frac{R{r}}{A}{B^\prime}_z)+\partial_r(r{{E^\prime}_{\phi^\prime}})-\partial_{\phi^\prime}(\frac{A}{R}{E^\prime}_r)=0\\
\partial_{t^\prime} {B^\prime}_{\phi^\prime} -\partial_r(\frac{A}{R}{E^\prime}_z)+\partial_z(\frac{A}{R}{E^\prime}_r)=0\\
\frac{\tilde {A}}{A}\partial_{t^\prime} {E^\prime}_z+\frac{r{R}}{A}\partial_{t^\prime} {B^\prime}_r+\frac{A}{R}\partial_{\phi^\prime}{{E^\prime}_z}-\partial_z(r{{E^\prime}_{\phi^\prime}})=0
\end{eqnarray}
These equations are different in form from those obtained in the non-rotating inertial frame which are given by the above equations with 
$A=R$ and $\tilde A = 0$. On the other hand in the limit $\beta \ll 1$, where  MFT reduces to GRT, from  (\ref{rot17}) we 
have $A\approx R$ and $\tilde A \approx 0$, i.e. the above homogeneous equations retain their inertial forms. In other words for points close to the rotation axis, where  MFT reduces to GRT, the {\it homogeneous} Maxwell equations are form invariant under GRT, a result first shown by Schiff \cite{Schiff}.\\
The same results as above could also be obtained by first writing the Maxwell equations in the non-rotating inertial frame using the field tensor in flat spacetime in cylindrical coordinates as follows: 
\begin{eqnarray}\label{rot21}
         {F}_{ij}=\left(
                  \begin{array}{cccc}
                    0 & -{E}_r & -{r}{E}_\phi & -{E}_z \\
                    {E}_r & 0 & {r}{B}_z & -{B}_\phi \\
                    {r}{E}_\phi & -{r}{B}_z & 0 & {r}{B}_r \\
                    {E}_z & {B}_\phi & -{r}{B}_r & 0 \\
                  \end{array}
                \right)
\end{eqnarray}
and then employ the general relation between the field tensors in the two frames,
\begin{equation}\label{rot22}
{F}_{ij}=\frac{\partial{{x^\prime}^m}}{\partial{{x}}^i}\frac{\partial{{x^\prime}^n}}{\partial{{x}}^j}{F^\prime}_{mn}
\end{equation}
to relate the primed and unprimed electromagnetic fields and finally replace the unprimed quantities (including partial differentials using Eq. (\ref{rot12})) by the primed ones.
So in general neither wave equation nor the Maxwell equations are invariant under MFT.
\section{Discussion and summary} 
Galilean rotational transformation is only true for centrally rotating observers. To relate the observations of inertial observers to those of eccentric, non-inertial ones (at large radii leading to relativist rotational velocities) on a rigidly rotating disk, one should either apply instantaneous LTs, introduced by Mashhoon et al. \cite {Mash0} in the context of hypothesis of locality, or alternatively look for consistent RRTs.
In the present article, we have discussed characteristics of a proposed RRT, dubbed as Franklin transformation, which relates coordinates of an event in two frames, one an inertial non-rotating frame and the other one rotating around their common axis with constant angular velocity $\Omega$ (measured by the inertial observers). The advantages and also drawbacks of this transformation specially with respect to the spacetime metric from the rotating observer's point of view as well as of its kinematical interpretation are pointed out. By introducing non-inertial observers at nonzero radii we have modified FT and showed how the modified transformation gives rise to a more consistent spacetime metric for these observers. The resulting spacetime metric includes two parameters, $\Omega$ and $R$, corresponding to the rotational angular velocity and radial position of these observers. Though a flat spacetime, it has a non-Euclidean spatial line element (found through $1+3$ formulation of spacetime decomposition) leading to non-Euclidean value for the circumference of a rotating disk or any other circle of a given radius.   In our setting of the problem of relativistic rotational transformations, there are three different kinds of observers: \textbf{I}- inertial non-rotating (laboratory) observers; \textbf{II}- centrally rotating (spinning) observer, and \textbf{III}- non-inertial rotating observers at nonzero radii (eccentric observers) who are rotating analogs of Rindler observers. In brief, following are the important features of the MFT:\\
{\bf 1}-Unlike FT it leads to a spacetime metric in the rotating frame which reduces to the spacetime metric obtained through GRT in the corresponding limit (i.e. close to the rotation axis).\\
{\bf 2}-Unlike in FT, the spacetime metric obtained via MFT preserves the {\it temporal} and {\it angular} isometries present in (\ref{rot3}).\\
{\bf 3}-At $R=0$ it reduces to the exact GRT as expected from its interpretation.\\
{\bf 4}-It gives a possible answer to the question: what is the spacetime metric for an eccentric observer on a rotating disk?\\
{\bf 5}-Related to the above point, at the position of an eccentric observer (i.e. at $r=R$), the spacetime metric is found to be form invariant (i.e. it reduces to the spatially Euclidean flat metric), a fact hinting toward a possible relation with Fermi metric and Fermi coordinates.\\
The last point above seems to be  interesting evidence reinforcing our interpretation of the MFT and its corresponding metric. The fact that the MFT includes two parameters, $\Omega$ and $R$, does not change its group character inherited from FT. Indeed, comparing (\ref{rot810}) and (\ref{rot811}), the group parameter in MFT is $v=c \tanh \beta $  whereas in FT it is $\Omega$ \footnote{As a matter of fact one could think of $R \phi$ in MFT, or $r\phi$ in FT, as an arclength coordinate.}. Indeed when it is compared to the coordinate transformation obtained in the approach based on the hypothesis of locality and Fermi metric (restricted to uniformly rotating observer) the appearance of $R$ is expected naturally as in that case the radial position of the eccentric observer enters the transformation both explicitly and also through the parameter $\beta$ (refer to \cite{Mash0}). The above-mentiond relation could be further investigated by a comparative study between the approaches based on MFT and its corresponding metric (\ref{rot11}) on the one hand, and the Fermi metric \cite{MTW} attributed to a uniformly rotating observer \cite{Mash0}, on the other hand. These are discussed and analyzed in detail in \cite{Nouri1}.\\
It is also shown explicitly that, against the previous claims, neither the Maxwell equations are invariant under FT or MFT, nor is the wave equation.\\
From the experimental and observational points of view it is expected that application of a relativistic rotational transformation to known physical effects related to the rotating systems and phenomena should lead to predictions different from those obtained through application of GRT or rotational transformations based on the hypothesis of locality. Some of the examples include transverse Doppler effect, Sagnac effect \cite{Sagna} and rotational properties of pulsars \cite{Kichen}. For a light source circling a receiver on a rotating disk, transverse Doppler effect will be affected naturally by FT and MFT, due to the nonlinear velocity distribution (\ref{rot7}) introduced in FT and this could be the most feasible test of the validity of MFT. Also it is expected that employing a relativistic rotational transformation will lead to a relativistic Sagnac effect distinct from the one due to propagation of light in a non-vacuum medium where relativistic velocity addition rule applies. Finally, fastest rotating celestial objects (apart from the supermassive black holes) are pulsars and the fastest pulsar, named PSR J1748-2446ad, is located some 28,000 light-years from Earth in the constellation Sagittarius and is spinning at 716 Hertz. If its radius is taken to be 16 km it will have a Galilean linear velocity of 75000k km/s, i.e. about $\% 25$ that of light speed at the equator. It is expected that at this rotational velocity a relativistic rotational transformation is at work and observationally effective. To look for experimental signatures of departure from GRT or rotational transformations based on hypothesis of locality, other physical effects (mainly electromagnetic in nature) which have already been studied in rotating frames \cite {Mash1,Mash2} should be reconsidered and interpreted in terms of MFT. In this regard, some of the rotational phenomena mentioned above are studied comparatively using both MFT and the formalism based on hypothesis of locality in \cite{Nouri1}.\\
Another very interesting issue which needs a careful treatment is the Unruh effect for uniformly rotating eccentric observers, which is already expected to be a controversial issue. By the above discussions it seems inevitable that one should employ a relativistic rotational transformation, such as MFT, to see whether eccentric observers detect any particle in the vacuum state of an inertial observer. These matters will be discussed elsewhere \cite{Nouri2}.
\section *{Acknowledgments}
The authors would like to thank University of Tehran for supporting this project under the grant no. 6101040/1/5 provided by the research council. 
\appendix
\section{$1+3$ (threading) formulation of spacetime decomposition and spatial distance}\label{Appa}
To define spatial metric and spatial distances in a given spacetime (metric) one could choose different spacetime decomposition 
formalisms. In our study we have  employed the $1+3$ (or threading) formulation of spacetime decomposition. Unlike the $3+1$ (or foliation) formulation 
of spacetime decomposition \cite{MTW} in which spacetime is foliated into constant-time hypersurfaces, in the $1+3$ formulation it is 
decomposed into threads tracking history of each spatial point.
This formulation of spacetime decomposition starts from the following form for the metric of a {\it stationary} spacetime $(\mathcal{M}, g_{ab})$ \cite{Landau},
\begin{eqnarray}\label{appa1}
ds^2 = d\tau_{syn}^2 - dl^2 = g_{00}(dx^0 -{{A_g}_\alpha}dx^\alpha)^2 -\gamma_{\alpha\beta}dx^{\alpha}dx^{\beta},\;\;\;\alpha, \beta = 1,2,3
\end{eqnarray} 
in which all the metric components are time-independent, i.e. the coordinate system is adapted to the timelike Killing vector field of the spacetime
($\xi^a\doteq\delta_0^a=(1,0,0,0)$). Also $d\tau_{syn}=\sqrt{g_{00}}(dx^0 - {A_g}_{\alpha} dx^{\alpha}) $ is the synchronized proper time,  ${A_g}_{\alpha}=-\frac{g_{0\alpha}}{g_{00}}$ is the so-called gravitomagnetic potential, and
\begin{eqnarray}\label{appa2}
dl^2 = \gamma_{\alpha\beta}dx^{\alpha}dx^{\beta} = (-g_{\alpha\beta} + 
\frac{g_{0\alpha}g_{0\beta}}{g_{00}})dx^\alpha dx^\beta
\end{eqnarray}
is the spatial line element (also-called the radar distance element) of the 3-space (denoted by $\Sigma_3$) in terms of its three-dimensional spatial metric $\gamma_{\alpha\beta}$. It should be noted that the 3-space $\Sigma_3$, introduced in this formalism, is the quotient space/manifold $\frac{\cal M}{G_1}$ where $G_1$ is the one-dimensional group of motions generated by the timelike Killing vector field of the underlying spacetime \cite{Geroch,exact}. It should be noted that $\Sigma_3$ is a manifold but not necessarily a submanifold (hypersurface) of the original spacetime manifold ($\cal M$).
Indeed it is the integral of the above line element which gives the spatial distance between two events with spatial 
coordinates $x_i^\alpha$ and $x_f^\alpha$ \cite{Landau},
\begin{eqnarray}\label{appa3a}
L =\int_{x_i^\alpha}^{x_f^\alpha} dl 
\end{eqnarray} 
For two simultaneous events at nearby points $x^\alpha$ and $x^\alpha + dx^\alpha$ the difference between their coordinate (world) time is given by
\begin{eqnarray}\label{appa3}
\Delta x^0 = A_\alpha dx^\alpha,
\end{eqnarray} 
This allows one to synchronize clocks in an infinitesimal region of space and also along any open curve. But
synchronization of clocks along a closed path is generally not possible, since upon returning to the initial point 
the world time difference is not zero and in the case of {\it stationary} spacetimes is given by the line integral
\begin{eqnarray}\label{appa4}
\Delta x^0 = \oint A_\alpha dx^\alpha
\end{eqnarray} 
taken along the closed path.
Using the above equation the world-time difference for two photons starting at the same point but traveling in opposite directions (clockwise and counter clockwise) along a circle of radius $R$ on a disk rotating with angular velocity $\Omega$ such that $\frac{\Omega R}{c} \ll 1$ is given by
\begin{eqnarray}\label{appa5}
\Delta t = 4\pi {R}^2 \frac{\Omega}{c^2} 
\end{eqnarray}
This difference, which leads to a phase shift $\delta \phi = \frac{2\pi c \Delta t}{\lambda}$ could also be obtained through classical reasoning by an inertial non-rotating observer and is the theoretical basis of the so-called Sagnac effect \cite{Sagna} or in its modern version, ring laser interferometry.\\
The 3-velocity of a test particle is defined in terms of the synchronized proper time as follows:
\begin{eqnarray}\label{appa41}
v^\alpha = \frac{dx^\alpha}{d\tau_{syn}} = \frac{c dx^\alpha}{\sqrt{g_{00}}(dx^0 - {A_\alpha} dx^\alpha)},
\end{eqnarray} 
where now using (\ref{appa1}) and (\ref{appa41}) the spacetime line element could be written as follows:
\begin{eqnarray}\label{appa51}
ds^2 = c^2 d\tau_{syn}^2 (1-\frac{v^2}{c^2}).
\end{eqnarray}
Now the components of the 4-velocity $u^i = \frac{dx^i}{ds} \;\;\;\; (i=0,1,2,3)$, in terms of the components of the 3-velocity are given by
\begin{eqnarray}\label{appa61}
u^0 =\frac{1}{\sqrt{g_{00}}\sqrt{1-v^2/c^2}} + \frac{A_\alpha v^\alpha}{\sqrt{1-v^2/c^2}}~~~;~~~u^\alpha = \frac{v^\alpha}{\sqrt{1-v^2/c^2}},
\end{eqnarray}
where in the comoving frame, $v^\alpha =0 $, it reduces to $u^i = (\frac{1}{\sqrt{g_{00}}}, 0, 0, 0)$ as expected. Also using the above definition of the 3-velocity one could show that, in a stationary spacetime, the energy of the particle defined as the time component of its 4-momentum is given by,
\begin{eqnarray}\label{appa62}
E \equiv P_0 = c g_{0i} u^i = \frac{mc^2 \sqrt{g_{00}}}{\sqrt{1-\frac{v^2}{c^2}}},
\end{eqnarray}
which is a conserved quantity reducing to $mc^2 \sqrt{g_{00}}$ in the comoving frame \cite{Landau}. It is this same formulation of spacetime decomposition which allows one to use analogy with electromagnetism  and define {\it gravitoelectric} and {\it gravitomagnetic} fields as follows:
\begin{eqnarray}
\textbf{E}_g=-\dfrac{\nabla h}{2h} ~~~\textbf{B}_g=\nabla \times \textbf{A}.
\end{eqnarray}
In terms of the above fields and in the context of the so-called {\it gravitoelectromagnetism}, vacuum Einstein field equations could be rewritten 
in the following {\it quasi-Maxwell} form \cite{Lynd,Nouri-Tavan},
\begin{gather}
\nabla \times ~\textbf{E}_g=0 \;\;\; ; ~~~\nabla \cdot  \textbf{B}_g=0 \label{r00} \\
\nabla \cdot \textbf{E}_g= 1/2 h B^2_g+E^2_g \label{r01} \\  
\nabla \times  (\sqrt{h}\textbf{B}_g)=2 \textbf{E}_g \times (\sqrt{h}\textbf{B}_g)] \label{r02} \\
{^{(\Sigma_3)}}R^{\mu\nu}=-{E}_g^{\mu;\nu} + \frac{1}{2}h(B_g^\mu B_g^\nu - B_g^2 \gamma^{\mu\nu})+ {E}_g^\mu E_g^\nu. \label{r03}
\end{gather}
where ${^{(\Sigma_3)}}R^{\mu\nu} $ is the three-dimensional Ricci tensor of the 3-space constructed from the three-dimensional metric $\gamma_{\alpha\beta}$ 
in the same way that the usual 4-dimensional Ricci tensor $R^{ab}$ is made out of $g_{ab}$. The first two equations ((\ref{r00})) are direct consequences of our definitions of gravitoelectric and gravitomagnetic fields and the original ten field equations are now given by those constituted in  (\ref{r01})-(\ref{r03}).\\
It should also be noted that in the above equations all the differential operations are defined in the 3-space with metric $\gamma_{\alpha\beta}$ \cite{Landau,Lynd}, in particular divergence and curl of a vector are defined as follows:
\begin{eqnarray}\label{appa9}
div\textbf{V}=\frac{1}{\sqrt{\gamma}}~\frac{\partial}{\partial{x^\alpha}}(\sqrt{\gamma}~V^\alpha)
~~~,~~~(curl\textbf{V})^\alpha = \frac{1}{2\sqrt{\gamma}}~\epsilon^{\alpha\beta\gamma}
(\frac{\partial{V_\gamma}}{\partial{x^\beta}}-\frac{\partial{V_\beta}}{\partial{x^\gamma}}),
\end{eqnarray}
in which $\gamma=det~\gamma_{\alpha\beta}$ and one can show that 
\begin{eqnarray}\label{appa10}
-g = h \gamma.
\end{eqnarray}
\pagebreak

\end{document}